\documentclass[journal, a4paper]{IEEEtran}
\usepackage[T1]{fontenc}
\usepackage[latin9]{inputenc}
\usepackage{color}
\usepackage{amsmath}
\usepackage{amsthm}
\usepackage{amssymb}
\usepackage{esint}
\usepackage{verbatim}
\usepackage{graphicx}
\usepackage{stfloats}
\usepackage{subfigure}
\usepackage{dsfont}
\usepackage{bm}
\usepackage{multicol}
\usepackage{multirow}
\usepackage{tabularray}
\usepackage{makecell}
\usepackage{algorithm}
\usepackage{algorithmic}
\usepackage{mathtools}
\usepackage{array}
\usepackage[numbers,sort&compress]{natbib}

\makeatletter

\usepackage[normalem]{ulem}

\newtheorem{assp}{Assumption}

\begin{document}
\title{ Semantic MIMO:  Revisiting Linear Precoding in the Generative AI Era}

\author{ Chunmei Xu$^{\dagger}$, Yi Ma and Rahim Tafazolli\\
	{\small 6GIC, Institute for Communication Systems, University of Surrey, Guildford, UK, GU2 7XH \\E-mails: (chunmei.xu, y.ma, r.tafazolli)@surrey.ac.uk.}\\
}
	
		\maketitle
		\thispagestyle{empty}
		
	\begin{abstract}
	This paper revisits linear precoding, namely  match-filter (MF) and zero-forcing (ZF), in a semantic multiple-input multiple-output (MIMO) system empowered by generative AI. 
	The aim is to examine whether interference, channel state information (CSI) accuracy, and scalability limitations in conventional MIMO systems remain critical. Theoretical analysis, which is based on the generative inference model and Lipschitz continuous assumptions, reveals reduced sensitivity to interference and channel imperfections, as well as  performance inferiority in high-SINR regimes compared to conventional MIMO systems.
	Simulation results validate the analysis and show that MF achieves semantic performance comparable to ZF under both perfect and imperfect CSI. 
	These findings suggest that semantic MIMO relaxes the needs for aggressive interference mitigation and highly accurate CSI, while improving scalability with reduced computational and implementation complexity.
	\end{abstract}
	 
	\section{Introduction}
	Semantic communications (SemCom) have recently emerged as a paradigm shift beyond conventional bit transmission, aiming to convey the semantic meaning of a source rather than ensuring exact bit-level reconstruction \cite{gunduz2022beyond}. 
	Generative artificial intelligence (AI) further empowers SemCom by endowing the receiver with strong inference capabilities through the learned generative priors  \cite{xu2025lightcom}. Such inference capability enables robust, high-fidelity  semantic reconstruction under low or uncertain signal-to-noise ratio (SNR) and signal-to-interference-plus-noise ratio (SINR) conditions \cite{xu2026uplink}. This inference-driven paradigm fundamentally reshapes reliability requirements, motivating a redesign of physical-layer transmission strategies, particularly multiple-input multiple-output (MIMO) techniques that constitute a cornerstone of modern wireless communications \cite{heath2018foundations}. 
	
	Conventional MIMO performance is critically limited by several bottlenecks \cite{love2008overview, larsson2014massive, lu2014overview}.  
	First, the achievable spatial multiplexing gains are primarily limited by interference, which intensifies with the growth in the number of served users and network densification, requiring sophisticated interference suppression strategies. 
	Second, achieving such gains relies heavily on accurate channel state information (CSI), which is often difficult to obtain due to estimation errors, feedback quantization and delays, and mobility-induced channel mismatch. Third, advanced MIMO approaches typically require computationally intensive operations, such as high-dimensional matrix inversion for linear precoding. Their complexity scales rapidly with the system dimension, raising significant scalability and implementation concerns.

	{The generative AI-empowered SemCom paradigm is potential to alleviate these conventional MIMO limitations by leveraging the inference capability.}  Recent studies have investigated semantic transmission over MIMO channels,  demonstrating promising semantic performance \cite{wang2022transformer, weng2024semantic,wu2024deep,liang2025vision}. However, existing works primarily focus on semantic coding schemes and end-to-end learning frameworks tailored for MIMO transmission. The impacts of interference and CSI imperfections on semantic MIMO systems remain largely unexplored. {Filling this gap is essential for guiding the design and optimization of semantic MIMO systems.}

	{This paper revisits classical linear precoding strategies in  semantic MIMO systems, where the generative AI models are deployed for inference-driven reconstruction.} We consider two representative precoding strategies, i.e., match-filter (MF) and zero-forcing (ZF), which adopts different treatments on interference.  Ignoring the interference, MF maximizes the desired signal power and offers low computational complexity. In contrast, ZF explicitly nulls interference by enforcing orthogonality among transmitted signals, but heavily relies on the CSI accuracy and incurs higher computational complexity due to matrix inversion operations.
	
	Simulation results demonstrate that semantic MIMO consistently outperforms conventional MIMO under low- or moderate-SNR regimes. Moreover, the MF precoding achieves semantic performance comparable to ZF precoding under both perfect and imperfect CSI conditions. This indicates that semantic MIMO significantly relaxes the needs for aggressive interference mitigation and highly-accurate CSI, while alleviating the scalability and implementation burdens  in the generative AI era.
 
	\section{System Model}
\begin{figure}[tp]
		\includegraphics[width=1\columnwidth]{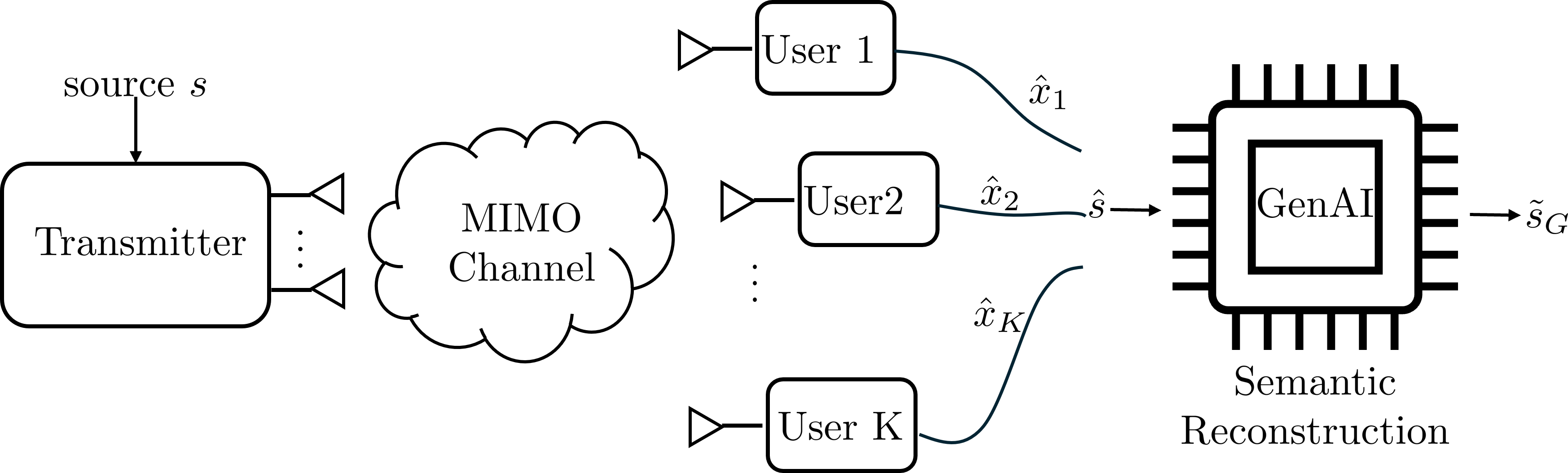}
		\caption{{The system model of the multi-user semantic MIMO system.}}
	\label{fig:system_model}
\end{figure}

	As illustrated in Fig.~\ref{fig:system_model}, we consider a multi-user semantic MIMO system, where a transmitter equipped with $N_t$ antennas serves $K$ single-antenna users. We adopt the inference-driven framework proposed in \cite{xu2026uplink} for image transmission, and exclude source coding to avoid confounding effects from compression distortion. 
	At the transmitter, the source $s$ is partitioned into $K$ bit sequences based on bit-level importance	\cite{xu2025dataimportanceJ} and then directly modulated using $M$-QAM into $K$ data streams $\{x_k\}_{k=1}^K$  \cite{xu2025lightcom}, with each transmitted to one user.
	The received $K$ data streams $\{\hat{x}_k\}_{k=1}^K$ are transferred to a compute-rich destination, where a large-scale generative model is employed for inference-driven semantic reconstruction. 
		
	\subsection{Channel Model}
	Denote the channel between the transmitter and the $k$-th user as $\mathbf{h}_k\in \mathbb{C}^{N_t \times 1}$, and the multi-user MIMO channel as the matrix 
	$\mathbf{H} = [\mathbf{h}_1, \cdots,\mathbf{h}_K ]$. $\mathbf{h}_k$ is modelled to follow the distribution:
	\begin{equation}\label{eq1}
		\mathbf{h}_{k} \sim \mathcal{CN}\!\left(0,\frac{1}{N_t}\mathbf I\right),
	\end{equation}where $\mathbf I$ is the {identity} matrix. The variance $1/N_t$ ensures that the average channel power is independent of the number of transmit antennas.

Obtaining perfect CSI is difficult due to practical factors such as estimation errors, feedback quantization and delays. By taking into account such channel imperfections, the channel can be further modelled as: 
	\begin{equation}\label{inaccurate channel}
		\mathbf{h}_k = \hat{\mathbf{h}}_k + \mathbf{e}_k,
	\end{equation}where $\hat{\mathbf{h}}_k$ is the CSI available at the transmitter, forming 	$\hat{\mathbf{H}} = [\hat{\mathbf{h}}_1, \cdots,\hat{\mathbf{h}}_K ]$. $\mathbf{e}_k$ represents the channel error. Assuming $\hat{\mathbf{h}}_k$ and  $\mathbf{e}_k$ are uncorrelated, the channel error $\mathbf{e}_k$ is modelled to following the distribution:
	\begin{equation}
		\mathbf{e}_{k} \sim \mathcal{CN}(0,\sigma_e^2\mathbf I),
	\end{equation}
	where the variance $\sigma_e^2$ captures the severity of CSI imperfections.

\subsection{{Signal Model}}
An equal power allocation strategy is applied at the transmitter. The received signal at the $k$-th user is expressed by:
 	\begin{equation}\label{eq01}
 		y_k  = \mathbf{h}_k^\mathrm{H}\mathbf{f}_k\sqrt{p}{x}_k + \sum_{j\neq k}\mathbf{h}_k^\mathrm{H}\mathbf{f}_j\sqrt{p}{x}_j + v_k,
 	\end{equation}where ${x}_k$ is normalized such that $\mathbb E[\vert x_k\vert^2]=1$. $p$ is the transmit power, and $\mathbf{f}_k \in \mathbb{C}^{N_t \times 1}$ denotes the precoding vector with unit power such that $\|\mathbf{f}_k\|^2 = 1$. ${v}_k$ is the additive noise following the distribution of $\mathcal{CN}(0, \sigma^2_n)$.  We denote the transmit SNR per user as  $\mathrm{SNR}=p/\sigma_n^2$. 
	
	The desired signal and interference power are given by $
 	P_k = p\vert\mathbf{h}_k^\mathrm{H} \mathbf{f}_k\vert^2$ and  ${I}_{k} =  \sum_{j\ne k} p\vert \mathbf{h}_k^\mathrm{H} \mathbf{f}_j\vert^2$, respectively. Given the imperfect CSI model in \eqref{inaccurate channel}, their expectations with respect to the channel error are derived as:
\begin{align}\label{E_desired}
	\mathbb{E}[P_k]= { p |\hat{\mathbf{h}}_k^{\mathrm H}\mathbf{f}_j |^2}
	+{p\sigma_e^2},
\end{align}

\begin{align}\label{E_interference}
	\mathbb{E}[I_k]
	= {\sum_{j\ne k} p |\hat{\mathbf{h}}_k^{\mathrm H}\mathbf{f}_j |^2}
	+{p(K-1)\sigma_e^2}.
\end{align}

The first term in the righthand side of \eqref{E_interference}, denoted as $I_k^{\mathrm{precode}}\triangleq\sum_{j\ne k} p|\hat{\mathbf{h}}_k^{\mathrm H}\mathbf{f}_j|^2$, reflects the interference caused by the applied precoder $\mathbf{F}\triangleq[\mathbf{f}_1, \cdots,\mathbf{f}_K]$. The second term, denoted as  $I_k^{\mathrm{error}}\triangleq {p(K-1)\sigma_e^2}$, captures the residual interference stemming from the channel error $\mathbf{E}\triangleq[\mathbf{e}_1, \cdots,\mathbf{e}_K]$.  Denoting $P_k^{\mathrm{precode}}= p|\hat{\mathbf{h}}_k^{\mathrm H}\mathbf{f}_j|^2$, the average SINR of the $k$-th user can be expressed by:
\begin{align}\label{sinr}
 \gamma_{k} &= \frac{\mathbb{E}[P_k] }{\mathbb{E}[I_k] +  \sigma^2_n} = \frac{P_k^{\mathrm{precode}} + p\sigma_e^2 }{I_k^{\mathrm{precode}} + I_k^{\mathrm{error}} +  \sigma^2_n}.
\end{align} 

The  bit error rate (BER), which is the key performance indicator of conventional bit-centric MIMO system, of the $k$-th data stream can be approximately derived as: 
\begin{equation}\label{eq:BER_SNR_Uncoded}
	\mathrm{BER}_k = \alpha \mathcal{Q}\left(\beta\sqrt{{\gamma}_k}\right),
\end{equation}where  $\mathcal{Q}(x)=\frac{1}{\sqrt{2\pi}}\int_{x}^{\infty}\mathrm{exp}(\frac{-t^2}{2})dt$ is the Gaussian Q-function. For $M$-QAM modulation, the parameters $\alpha$ and $\beta$ are given by $\alpha  =  \frac{4}{\log_2 M}(1-\frac{1}{\sqrt{M}})$ and $\quad \beta=\sqrt{\frac{3}{M-1}}$, respectively.

\subsection{Generative Inference Model}
At the destination, the received data streams are first combined such that $\hat{{s}}=\sum_{k=1}^K 2^{k-1}\hat{x}_k$ {due to the consideration of image sources}, which is a noisy version of $s$. It is subsequently processed by the generative AI model for semantic reconstruction. This generative inference process can be modeled, from an optimization perspective, as the solution to the following regularized inverse problem  \cite{Zhu_2023_CVPR}:
\begin{equation}
	\tilde{s}_G = G(\hat{s})
	= \arg\min_{\tilde{s}}
	\frac{1}{2}\Vert \tilde{s}-\hat{s}\Vert^2
	+ \lambda R_\Theta(\tilde{s}),
\end{equation}
where $R_\Theta(\tilde{s})=-\log P_\Theta(\tilde{s})$ with $P_\Theta(\tilde{s})$ representing the learned generative prior captured by the generative AI model, and $\lambda>0$ is the regularisation parameter. This formulation admits a maximum a posteriori (MAP) interpretation \cite{yip2007digital}.  

\begin{assp}
	Assume  the operator $G(\cdot)$ is Lipschitz continuous, satisfying the contraction property:
\begin{equation}\label{eq10}
	\Vert G(u)-G(v)\Vert \le \rho \Vert u-v \Vert,
\end{equation}where the factor $ \rho$ characterizes the generative inference capability. A smaller value of $\rho$ corresponds to stronger generative inference capability.
\end{assp}

Denote $\hat{s}_\epsilon$ as the received representation with a sufficiently small error such that $\epsilon=\mathbb E[\Vert \hat{s}_\epsilon - s\Vert]$. 
To account for potential generative prior mismatch, we consider semantic reconstruction bias and model it as:
\begin{equation}
	\mathbb E[\Vert G(\hat{s}_\epsilon)-s \Vert] \le \delta_\epsilon,
\end{equation}
where $\delta_\epsilon \ge 0$ reflects the generative inference bias associated with the error level $\epsilon$.

\section{Linear Precoding Revisit \label{Sec:III}}	
 This section details both MF and ZF precoding techniques and discuss their performance behaviour in  conventional MIMO systems. They are two representative precoding approaches that treat interference differently, which are designed based on  $\hat{\mathbf{H}}$ at the transmitter:

\subsubsection{MF Precoding} The MF precoding matrix, denoted as  $\mathbf{F}^{\mathrm{mf}} = [\mathbf{f}_1^{\mathrm{mf}}, \cdots, \mathbf{f}_K^{\mathrm{mf}}]$ aims to maximize the desired signal power $P_k$ irrespective of the interference $I_k$. It is designed as:
	\begin{equation}
	\mathbf{f}^{\mathrm{mf}}_k = \frac{\hat{\mathbf{h}}_k}{\Vert\hat{\mathbf{h}}_k\Vert}.
	\end{equation}Accordingly, we have:  
	
	\begin{align}\label{P_MF}
		  P_k^{\text{mf-precode}} =  p\Vert \hat{\mathbf{h}}_k\Vert^2,
	\end{align}
	
	\begin{align}\label{I_MF}
		 I_k^{\text{mf-precode}} =   p\sum_{j\ne k}    \frac{\vert\hat{\mathbf{h}}_k^\mathrm{H} \hat{\mathbf{h}}_j\vert^2}{\Vert\hat{\mathbf{h}}_j\Vert^2}\ge 0.
	\end{align}
	The interference caused by $\mathbf{F}$ exists even under perfect CSI. 
	
	The required computations only involve Hermitian transpose and vector normalization. The computational complexity is $\mathcal O(KN_t)$, which is linearly scaled with the number of users $K$ and the number of transmit antennas $N_t$.  

\subsubsection{ZF Precoding}
The ZF precoding matrix, denoted as $\mathbf{F}^{\mathrm{zf}} = [\mathbf{f}_1^{\mathrm{zf}}, \cdots, \mathbf{f}_K^{\mathrm{zf}}]$, is designed to eliminate multi-user interference by enforcing orthogonality among $K$ transmitted signals. 
Specifically, ZF precoding projects each signal onto the null space of the remaining users' channels. Denoting $
	 \hat{\mathbf{A}} = (\hat{\mathbf{H}}^{\mathrm{H}}\hat{\mathbf{H}})^{-1}$, the ZF precoding vector for the $k$-th user is given by:

\begin{equation}
		\mathbf{f}^{\mathrm{zf}}_k =   \frac{\hat{\mathbf{H}}\hat{\mathbf{a}}_k  }{\Vert \hat{\mathbf{H}}\hat{\mathbf{a}}_k \Vert},
\end{equation}where $\hat{\mathbf{a}}_k$ is the $k$-th column of $\hat{\mathbf{A}}$.  It satisfies $\hat{\mathbf{h}}_i^{\mathrm{H}} \mathbf{f}_k^{\mathrm{zf}} = 0, \forall i \neq k$.  Accordingly, we have:

\begin{equation}\label{P_ZF}
	 P_k^{\text{zf-precode}} =  p\vert\hat{\mathbf{h}}_k^\mathrm{H}  {\mathbf{f}}_k^{\mathrm{zf}}\vert^2\le P_k^{\text{mf-precode}},\end{equation}

\begin{equation}\label{I_ZF}
	I_k^{\text{zf-precode}} =   p\sum_{j\ne k}     \vert\hat{\mathbf{h}}_k^\mathrm{H}  {\mathbf{f}}_j^{\mathrm{zf}}\vert^2 = 0 \le I_k^{\text{mf-precode}},
\end{equation}where both equalites hold if and only if $\hat{\mathbf{h}}_i \perp  \hat{\mathbf{h}}_j, \forall i\neq j$. The interference component $I_k^{\text{zf-precode}}$ is always zero.   

The computational complexity of ZF precoding is dominated by the inversion of the $K \times K$ Gram matrix $\hat{\mathbf{H}}^{\mathrm{H}}\hat{\mathbf{H}}$, which incurs a cubic complexity of $\mathcal{O}(K^3)$. The formation of the Gram matrix and the vector normalization require additional $\mathcal{O}(K^2N_t)$ and $\mathcal{O}(KN_t)$, respectively. 
Therefore, the overall computational complexity is $\mathcal{O}(K^3 + K^2N_t + KN_t)$ \cite{lu2014overview}, which scales cubically with the number of users $K$ and linearly with the number of transmit antennas $N_t$.

As indicated by \eqref{eq:BER_SNR_Uncoded},  the BER performance is primarily limited by the noise in low-SNR regimes and by interference power in high-SNR regimes. Consequently, MF slightly outperforms ZF at low SNRs due to higher desired power, whereas its performance degrades severely at high SNRs due to higher precoder-induced interference. Moreover, the BER performance highly relies on the accuracy of the available CSI, particularly the  ZF precoding technique.   

\section{Theoretical Analysis\label{sec:IV}}

\begin{figure*}[thp]
	\centering
	\subfigure{
		\includegraphics[width=0.8\columnwidth]{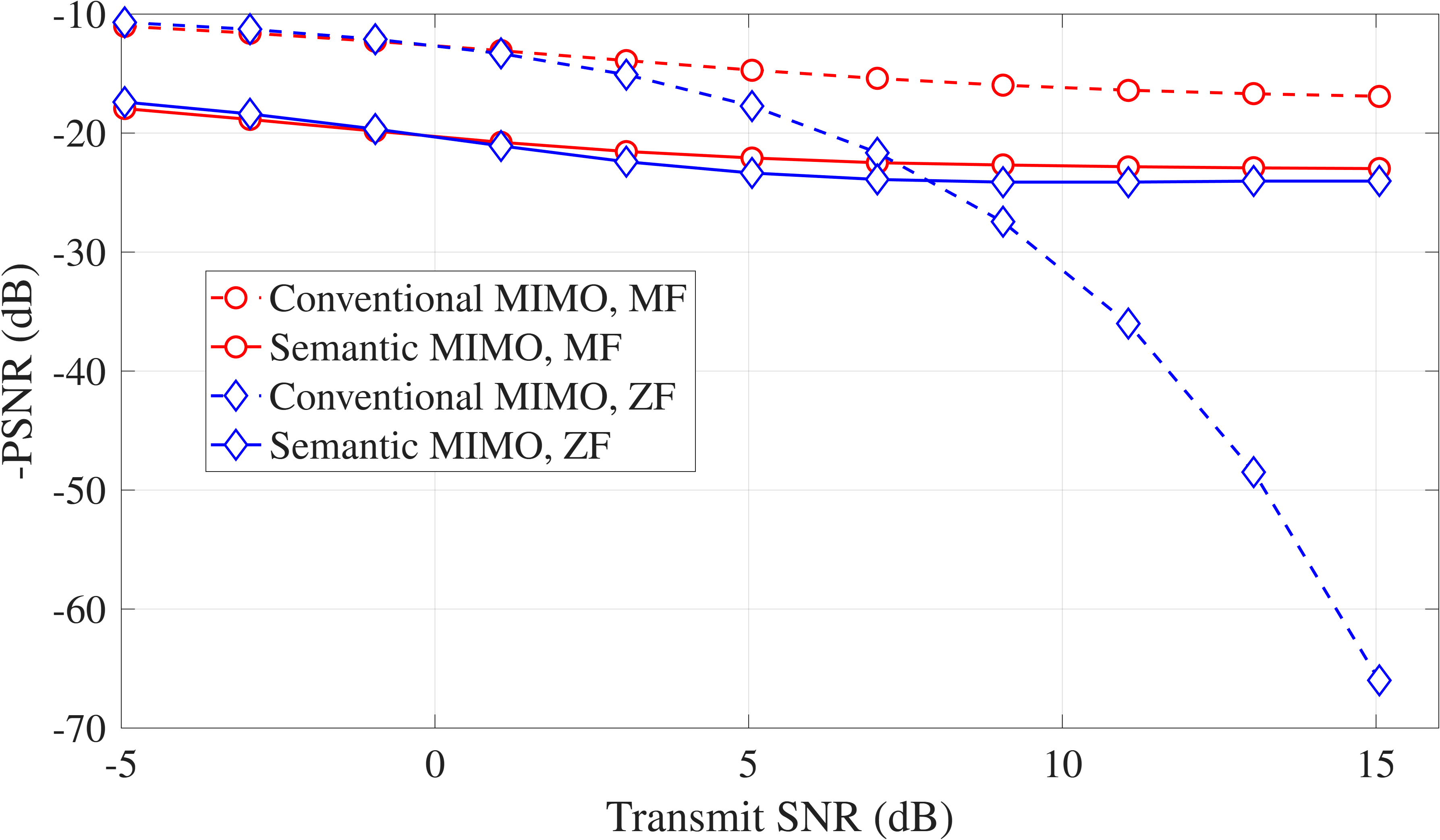}} \hspace{30pt}
	\subfigure{
		\includegraphics[width=0.8\columnwidth]{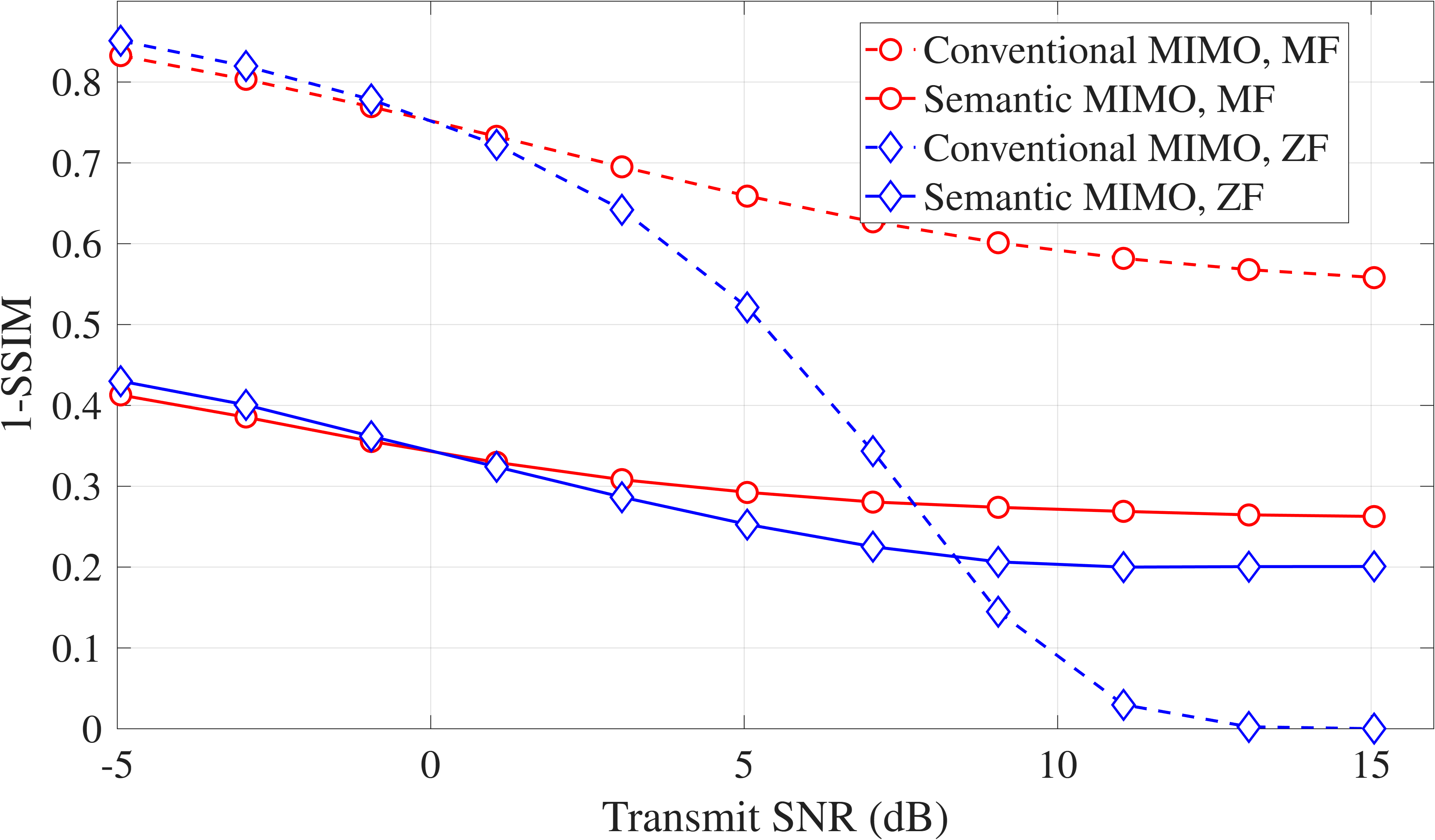}}
	
	\subfigure{
		\includegraphics[width=0.8\columnwidth]{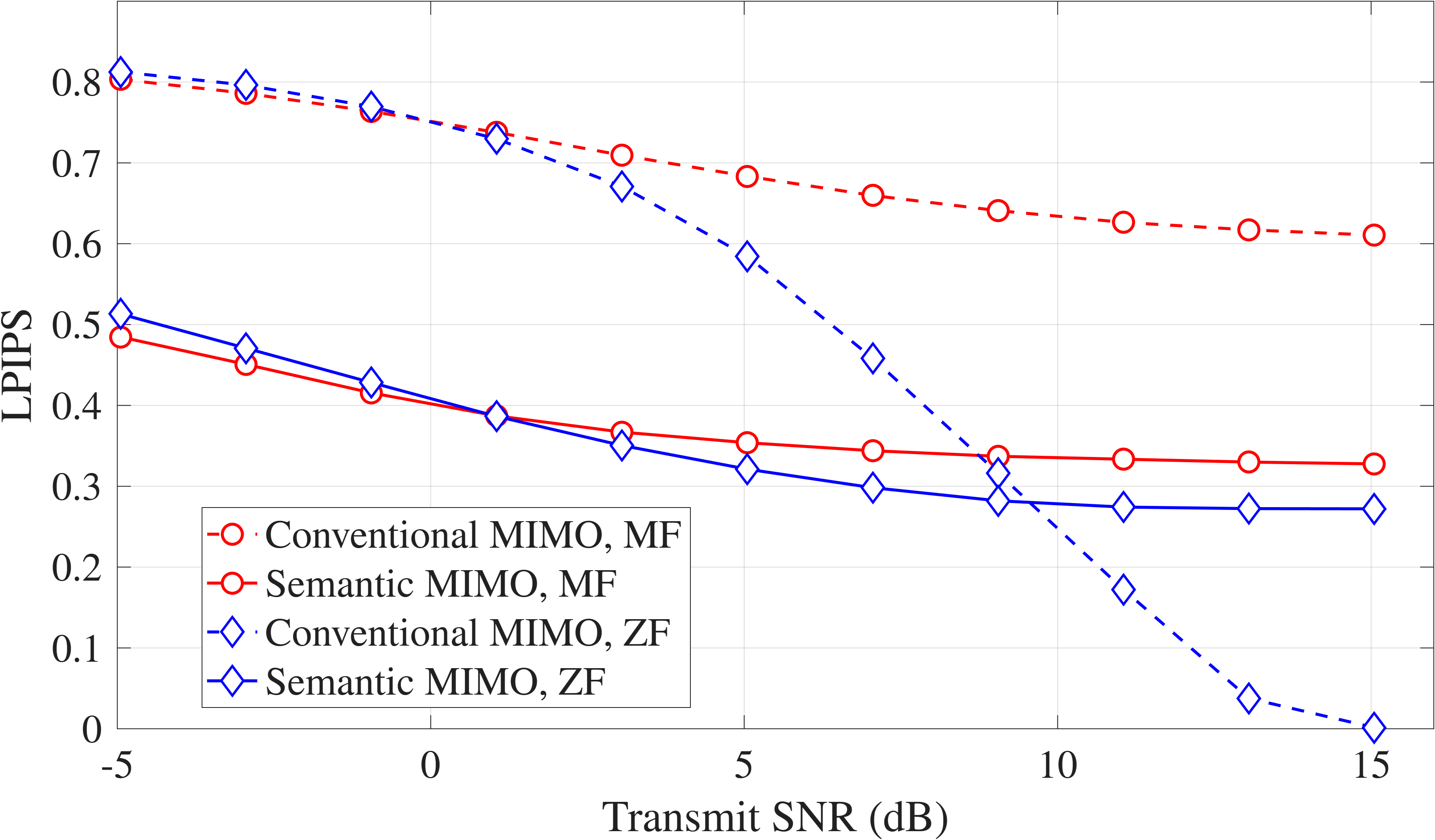}}  \hspace{30pt}
	\subfigure{
		\includegraphics[width=0.8\columnwidth]{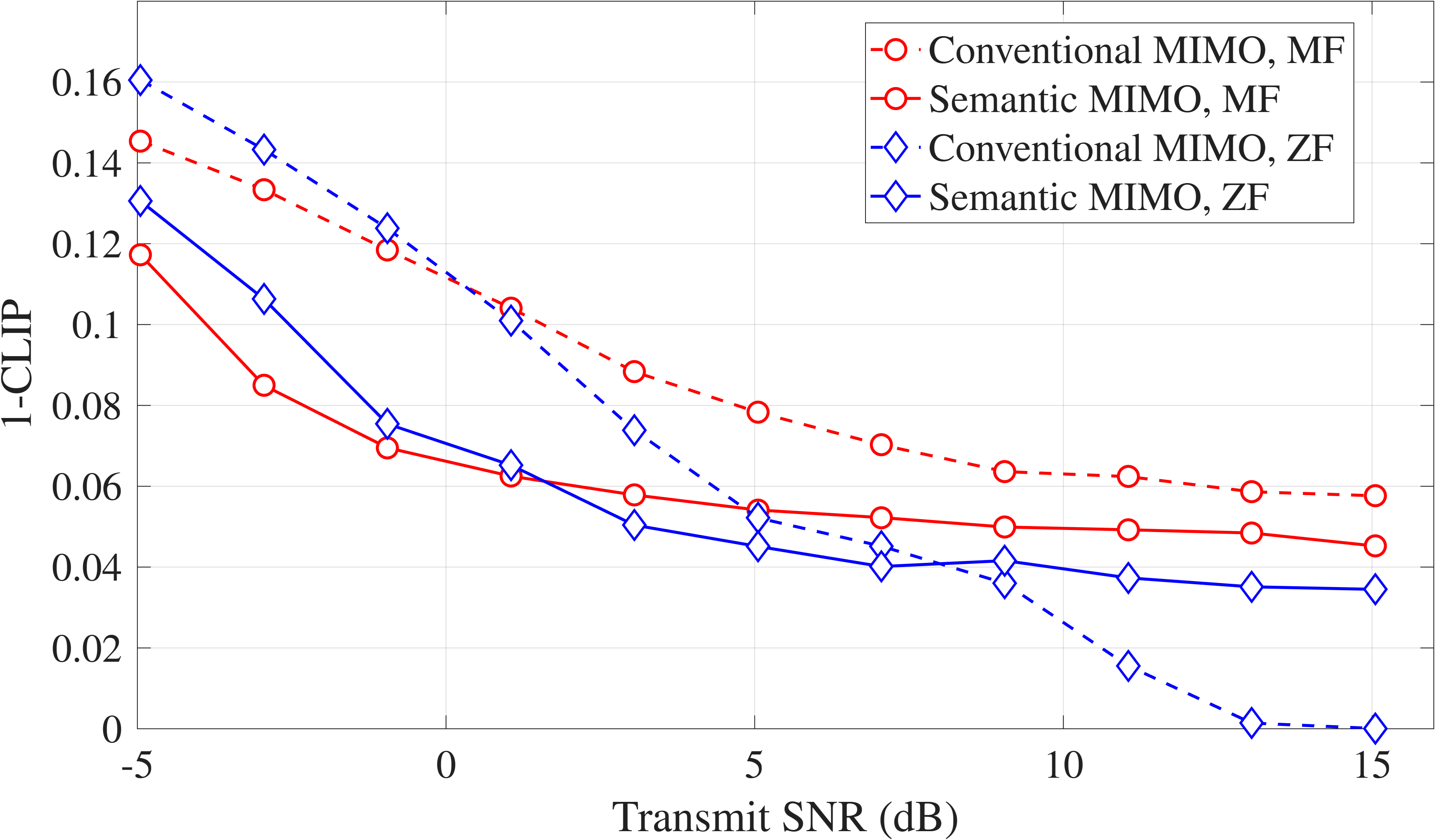}}
	\caption{Perfect CSI case: Semantic performance comparison across different transmit SNR per data stream.}
	\label{fig:results_11}
\end{figure*}
Consider a general semantic performance metric $\mathcal{M}(\tilde{s},s)$, where lower values indicate better semantic fidelity. Due to the nonlinear and data-driven nature of generative inference, deriving an explicit closed-form mapping between semantic performance and the received SINR is analytically intractable.  To enable tractable analysis, we make the following assumption:
\begin{assp}Assume the metric $\mathcal{M}(\tilde{s},s)$ is $\ell_{\mathcal{M}}$-Lipschitz continuous with respect to its first argument, satisfying the contraction property:
	\begin{equation}
		\big|\mathcal{M}(u,s) - \mathcal{M}(v,s)\big|
		\le \ell_{\mathcal{M}} \Vert u-v \Vert.
	\end{equation} 
\end{assp}

Using the triangle inequality, we obtain the following based on the contraction property in \eqref{eq10}:
\begin{align}
	\Vert \tilde{s}_G - s\Vert &\le \Vert P(\hat{s})-P(\hat{s}_\epsilon)\Vert + 	\Vert P(\hat{s}_\epsilon)-s	\Vert \nonumber \\
	&\le \rho \Vert \hat{s}-\hat{s}_\epsilon \Vert + \delta_\epsilon \nonumber \\
	&\le \rho \left(\Vert \hat{s}-{s}  \Vert + \epsilon \right) + \delta_\epsilon.
\end{align}Under Assumption 2, the semantic performance bound can be yielded:
\begin{equation}\label{inference_ineq}
	\mathbb{E} [\mathcal{M}(\tilde{s}_G,s)]
	\le
	\mathcal{M}(s,s)
	+  \rho\ell_{\mathcal{M}} (\mathbb E [\Vert \hat{s}-{s}  \Vert] + \epsilon)+ \ell_{\mathcal{M}}\delta_\epsilon,
\end{equation}where $\mathbb E [\Vert \hat{s}-{s}  \Vert]$ under channel-uncoded transmission can be approximated by 
\begin{equation}\label{eq:MSE_BER}
	\mathbb E [\Vert \hat{s}-{s}  \Vert] \approx  \sum_{k} 2^{k-1}\mathrm{BER}_k,
\end{equation}by assuming that at most one bit per pixel is incorrectly received.
  
Next, we analyze the two performance properties of the semantic MIMO system.
\subsubsection{SINR Dependence}
Differentiating the metric with respect to SINR $\gamma_k$ yields:
\begin{equation}
	 \frac{\partial \mathbb{E} [\mathcal{M}(\tilde{s}_G,s)]}{\partial \gamma_{k}}  
	\lesssim - \rho\ell_{\mathcal{M}}
	\frac{2^{k-1}\alpha}{2\sqrt{2\pi}\beta}e^{\frac{-\beta^2\gamma_{k}}{2}}
	(\gamma_{k})^{-\frac{1}{2}}.
\end{equation}It demonstrates that sensitivity of semantic performance to SINR is attenuated by the inference factor $\rho$. Given the strong inference capability, i.e., $\rho \ll 1$, the semantic performance becomes robust to SINR variation, indicating strong tolerance to interference and channel imperfections. As a consequence, the considerable SINR gains achieved by ZF precoding over MF precoding translate into limited semantic performance improvement.  

\subsubsection{High-SINR Inferiority}
For conventional MIMO systems employing a non-generative reconstruction, specifically the identity mapping $\tilde{s}_0=\hat{s}$,  the corresponding semantic performance is bounded by:
\begin{equation}
	\mathbb{E} [\mathcal{M}(\tilde{s}_0,s)]
	\le
	\mathcal{M}(s,s)
	+\ell_{\mathcal{M}}\mathbb E [\Vert \hat{s}-{s}  \Vert].
\end{equation}

Comparing with the performance bound in \eqref{inference_ineq}, semantic MIMO with generative inference is advantageous approximately when:
\begin{equation}
\rho (\mathbb E [\Vert \hat{s}-{s}  \Vert] + \epsilon)+ \delta_\epsilon  < \mathbb E [\Vert \hat{s}-{s}  \Vert].
\end{equation}Equivalently, the inference reconstruction becomes inferior when:
\begin{equation}
	 \mathbb E [\Vert \hat{s}-{s}  \Vert] < \frac{\rho\epsilon}{1-\rho}+\frac{\delta_\epsilon}{1-\rho},
\end{equation}
which corresponds to sufficiently high-SINR regimes where interference is negligible and generative prior-induced bias dominates.

\section{Simulation Results and Discussions \label{Sec:IV}}
This section examines the semantic performance of MF and ZF precoding in the considered semantic MIMO system, and compares it with conventional MIMO. 

 {\it{Parameter setup:}}  At the destination with a cloud platform, the generative AI model, namely the SUPIR \cite{yu2024scaling}, is deployed for semantic-level reconstruction of image sources. The number of transmit antennas and users are set to $N_t=16$ and $K=8$, respectively. The QAM modulation with an order of $M=4$ is adopted. The noise variance is set to $\sigma_e^2=1$. {For images, PSNR measures pixel-wise fidelity, while SSIM captures structural similarity aligned with human perception. LPIPS evaluates perceptual similarity using deep feature embeddings, and CLIP-based metrics assess semantic consistency in a joint vision-language space.} For performance evaluation, we adopt -PSNR, 1-SSIM, LPIPS, and 1-CLIP as semantic performance metrics to ensure lower values with better semantic reconstruction quality.   Both perfect and imperfect CSI cases are considered, as detailed below.

	{\it{Case 1: Perfect CSI.}} In this case,  $\hat{\mathbf{H}}=\mathbf{H}$. As analyzed, the ZF precoding can ideally eliminate the multi-user interference, whereas the MF precoding suffers from residual interference $I_k^{\text{mf-precode}}$. 
 Fig. \ref{fig:results_11} shows that the semantic MIMO system outperforms conventional MIMO system when MF is applied. When applying ZF, conventional MIMO outperforms semantic MIMO  in high-SNR regime (i.e., $\mathrm{SNR}\ge 10$ dB), due to the elimination of interference. In contrast, under low- and moderate-SNR conditions that are more representative of practical wireless environments (i.e., $\mathrm{SNR} < 10$~dB), the semantic MIMO outperforms conventional MIMO.
	Furthermore, despite the presence of residual interference, the MF precoding achieves semantic performance comparable to that of ZF in semantic MIMO. Although the performance gap between MF and ZF slightly increases with SNR, it remains marginal compared to the substantial performance disparity observed in conventional bit-centric MIMO systems. 
	These results confirm the powerful inference capabilities of the semantic MIMO by leveraging generative AI, and indicate that semantic MIMO is inherently more tolerant to interference.

\begin{figure*}[tp]
		\centering
		\subfigure{
			\includegraphics[width=0.8\columnwidth]{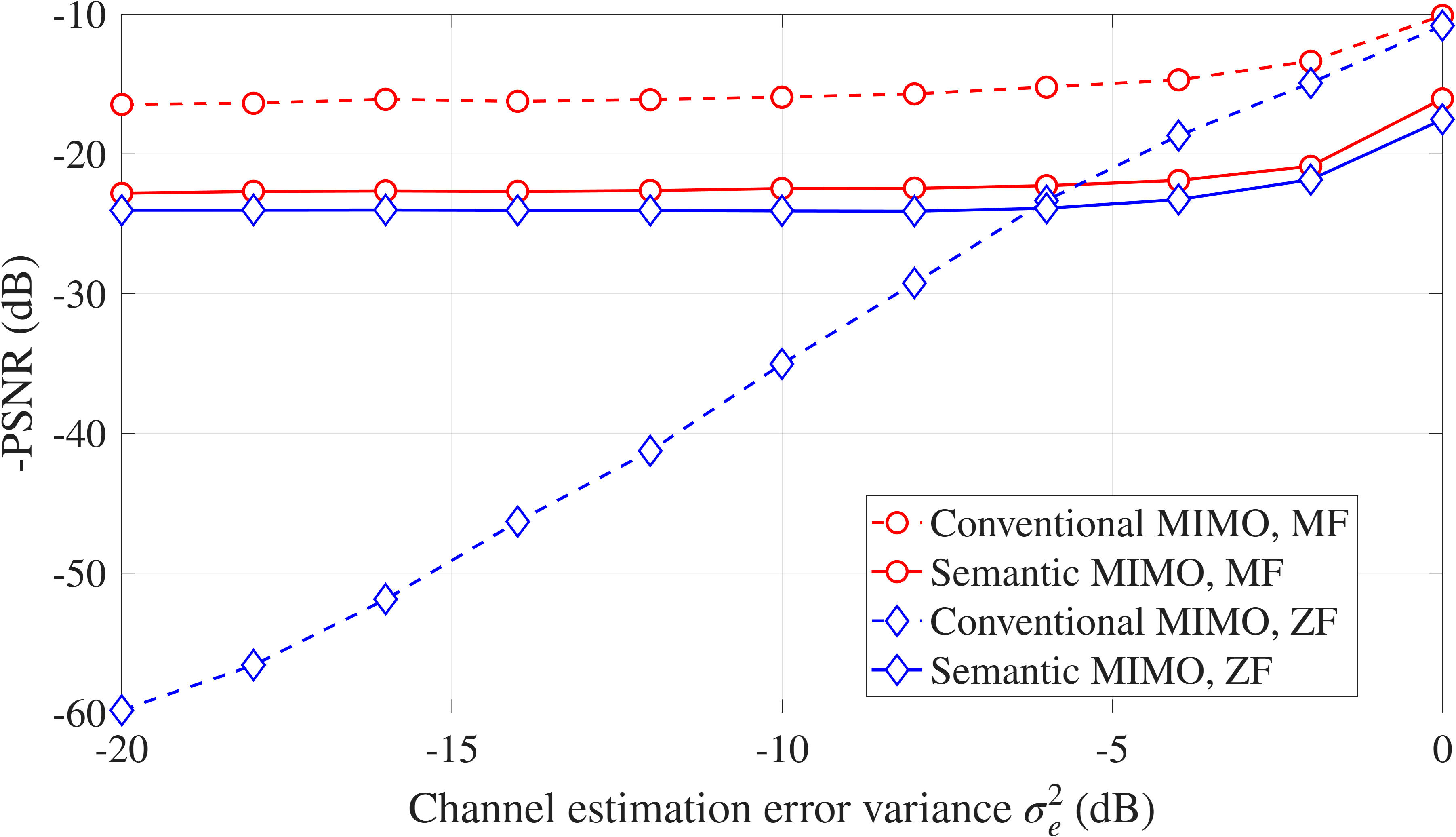}} \hspace{30pt}
		\subfigure{
			\includegraphics[width=0.8\columnwidth]{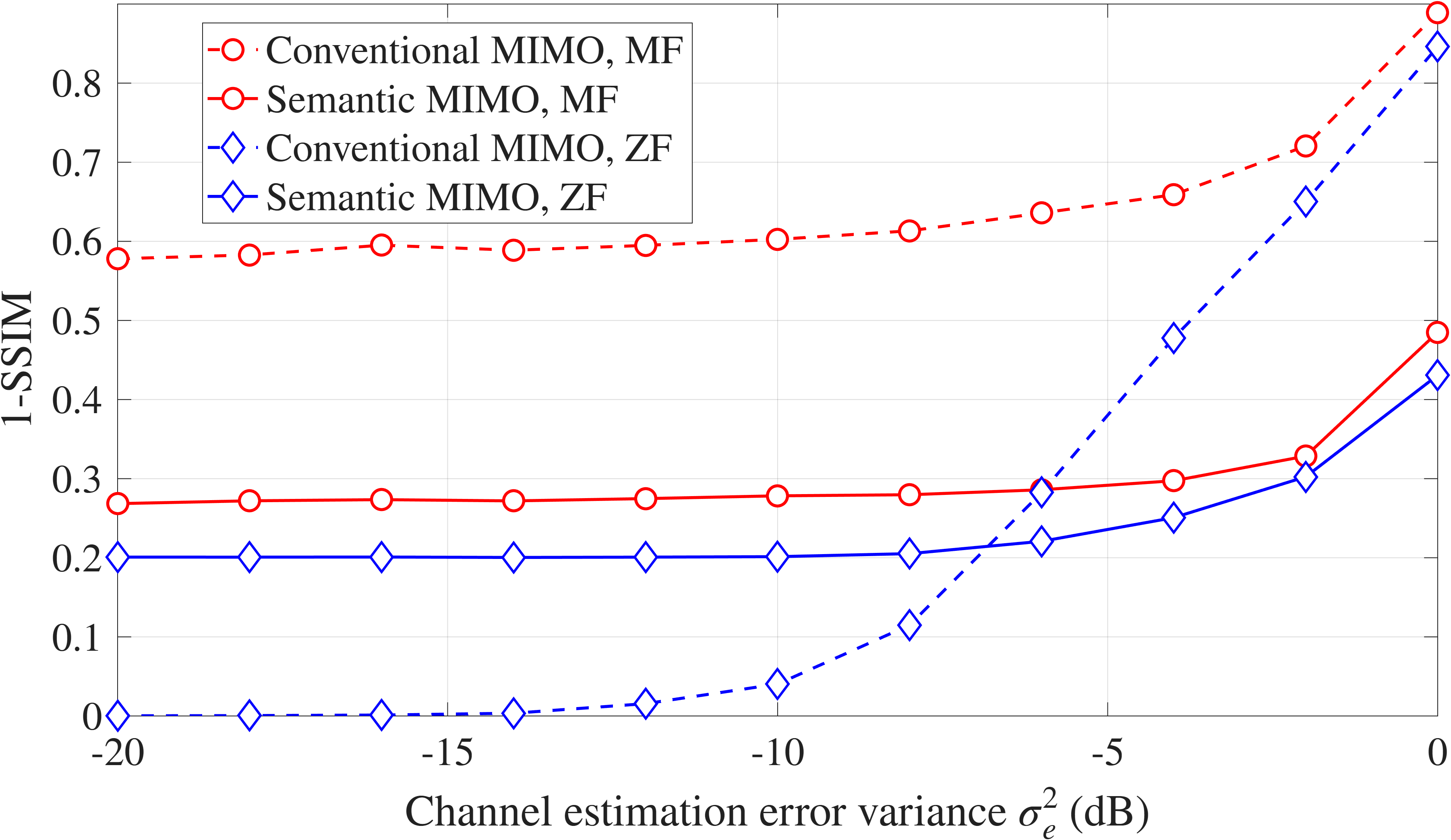}}
		\subfigure{
			\includegraphics[width=0.8\columnwidth]{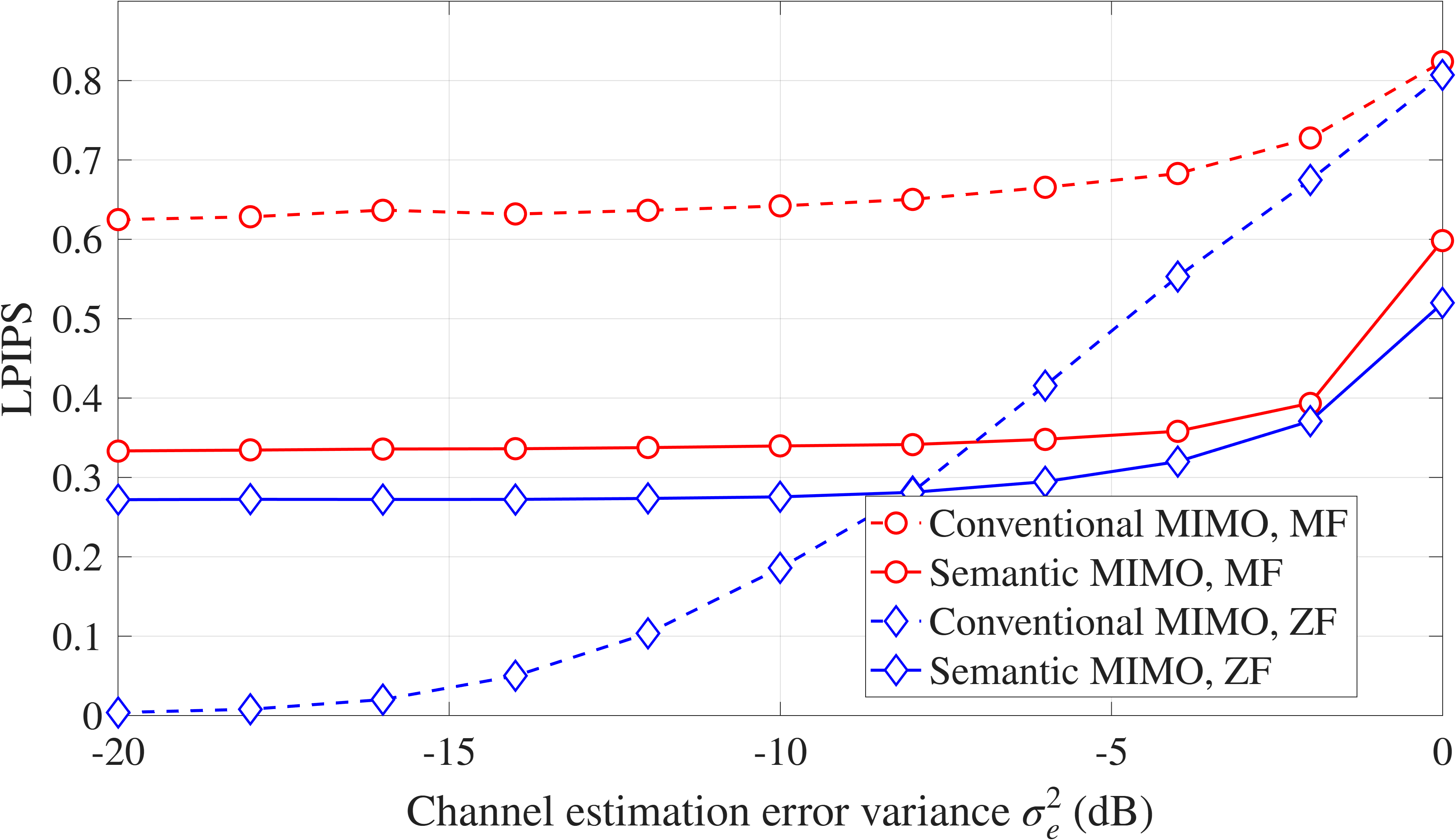}}  \hspace{30pt}
		\subfigure{
			\includegraphics[width=0.8\columnwidth]{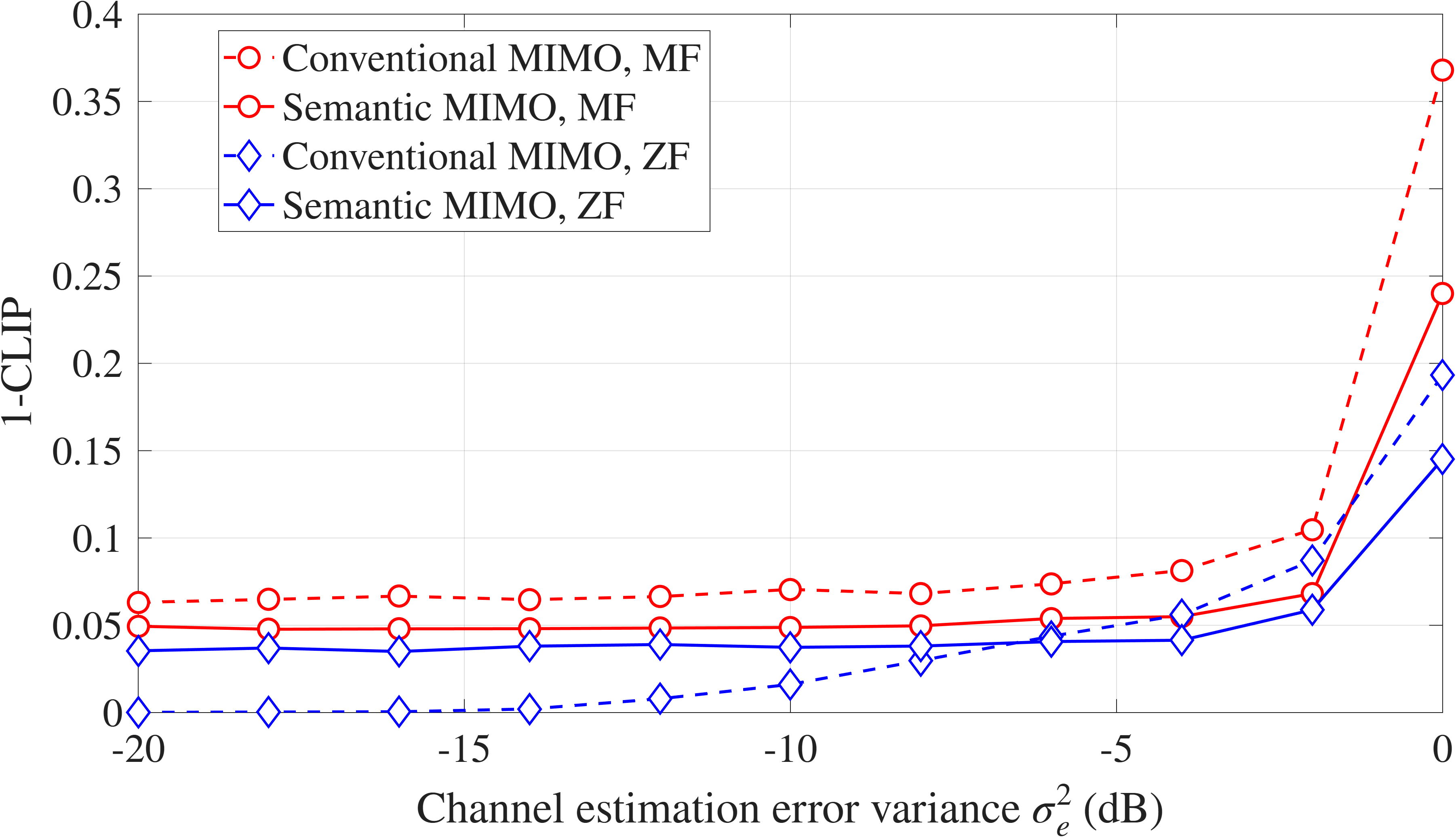}}
		\caption{Imperfect CSI case: Semantic performance comparisons between MF and ZF precoding techniques, where $\mathrm{SNR}=15$ dB.}
		\label{fig:results_21}
	\end{figure*}

	{\it{Case 2: Imperfect CSI.}} In this case, $\hat{\mathbf{H}}\neq\mathbf{H}$. We investigate the impact of channel errors, which induce interference  under both precoding schemes. 
	Fig. \ref{fig:results_21} provides the semantic performance comparisons. When MF precoding is applied, semantic MIMO consistently outperforms conventional MIMO across the entire range of channel error variance $\sigma_e^2$.  For ZF precoding,   conventional MIMO achieves better semantic performance only under small channel error variance (i.e., $\sigma_e^2\le -6$ dB), while semantic MIMO becomes increasingly advantageous as $\sigma_e^2$ grows. Furthermore, under semantic MIMO, ZF consistently achieves slightly better performance than MF. This can be attributed to the lower interference of ZF compared to MF, as reflected by the interference expressions in \eqref{I_MF} and \eqref{I_ZF}.
  	These results show that semantic MIMO with ZF precoding is less sensitive to channel errors than the conventional one, indicating relaxed requirements on the perfection of CSI.

\begin{figure*}[tp]
	\centering
	\subfigure[]{
		\includegraphics[width=0.95\columnwidth]{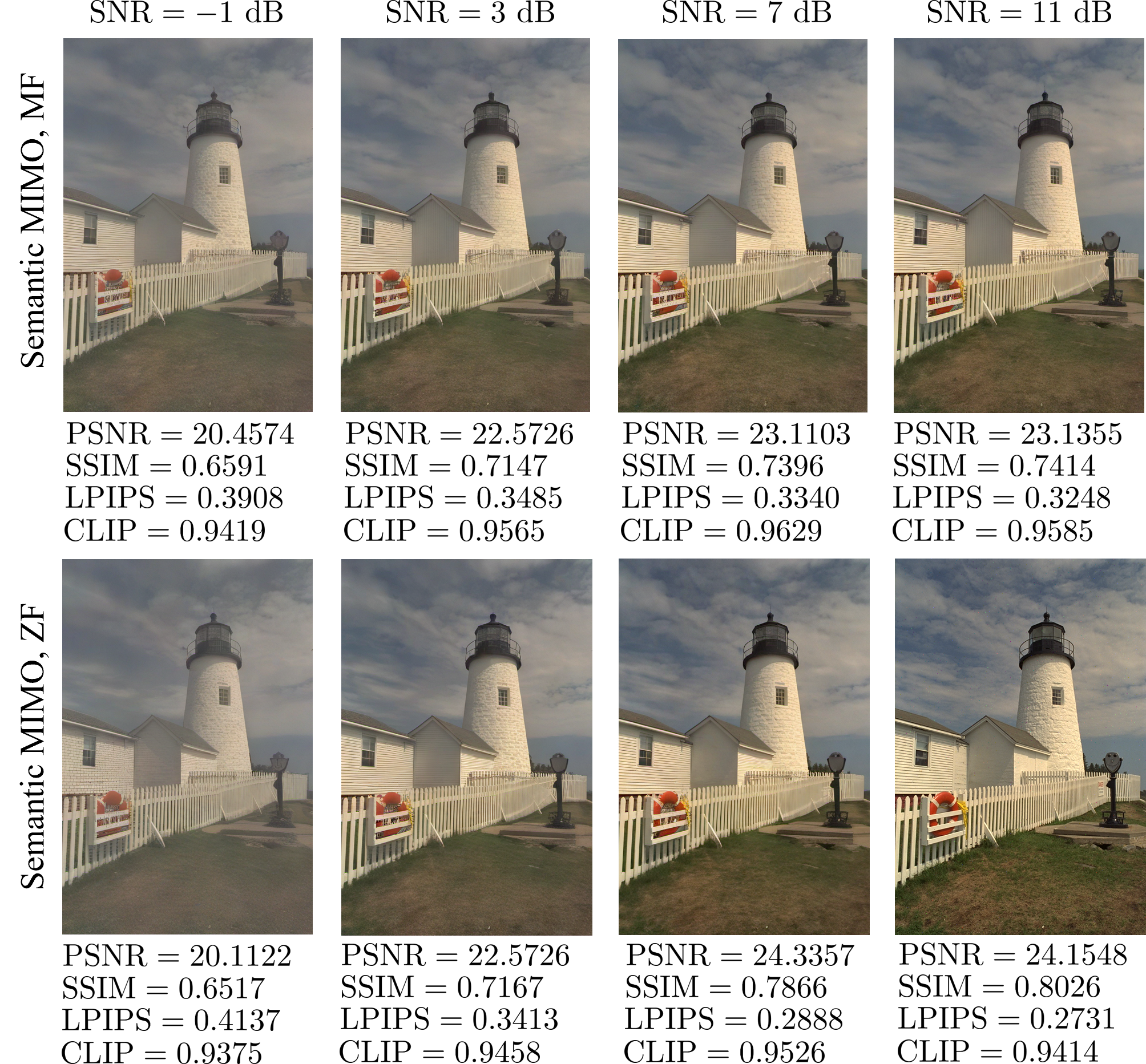}}  
	\subfigure[]{
		\includegraphics[width=0.95\columnwidth]{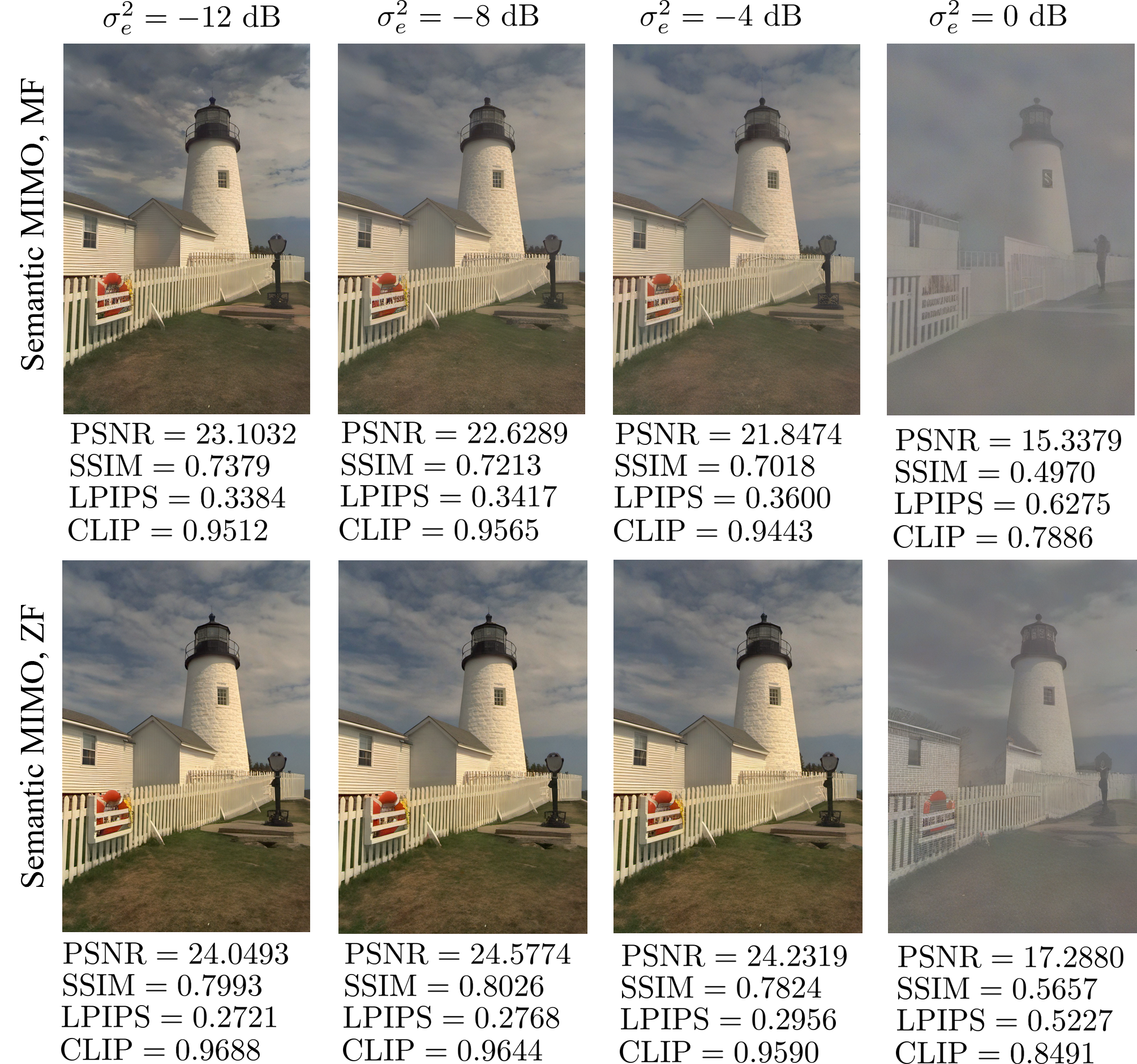}}   
	\caption{Visual illustration of reconstructions under (a) perfect CSI case, and (b) imperfect CSI case.   }
	\label{fig:results_3}
\end{figure*}

 Fig.~\ref{fig:results_3} presents the reconstructed images under the semantic MIMO system, providing a more intuitive visual comparison between the MF and ZF precoding schemes. These results  confirm our quantitative observations that MF achieves semantic reconstruction quality comparable to that of ZF. 

\section{Key Findings and Design Insights}
The simulation results in Sec. \ref{Sec:IV} confirm the performance analysis in Sec. \ref{Sec:IV}, and reveal several important findings and insights: 
\begin{itemize}
	\item[1)] \textbf{Improved semantic performance by inference capability:}  Semantic MIMO empowered by generative AI exhibits fundamentally different performance behaviour from the conventional MIMO system. By leveraging powerful inference capabilities, semantic MIMO consistently achieves superior semantic performance compared to conventional MIMO under practical wireless environments. This highlights the strong potential of inference-driven communication frameworks to improve semantic performance in the generative AI era.

	\item[2)] \textbf{Relaxed interference suppression need:} 
	Under perfect CSI conditions, MF precoding achieves semantic performance comparable to ZF, despite its limited interference suppression capability. This observation indicates that semantic MIMO is more tolerant to interference in addition to noise,  thereby	largely relaxing the need for aggressive interference mitigation required in conventional MIMO systems. 
	
	\item[3)] \textbf{Reduced reliance on accurate CSI:} Under imperfect CSI conditions, both MF and ZF achieve comparable semantic performance across a wide range of channel estimation error variances. This demonstrates that semantic MIMO fundamentally reduces the reliance on instantaneous accurate CSI. As a result, pilot overhead and CSI feedback requirements can be substantially relaxed particularly for massive MIMO configurations, thereby saving valuable system resources that can be exploited to further enhance communication efficiency.

	\item [4)] \textbf{Improved computational and implementation scalability:} 
	 MF precoding incurs significantly lower computational cost than ZF, while providing comparable semantic performance.   This indicates that the conventionally critical scalability bottleneck can be effectively alleviated in semantic MIMO systems by adopting lightweight precoding strategies such as MF. Moreover, a more practical codebook-based beamforming technique can achieve competitive semantic performance while significantly reducing implementation complexity.	This enables practical implementation in large-scale deployment.
\end{itemize}

	\section{Conclusion}
  	This paper revisited linear precoding in a generative AI-enabled semantic MIMO system to examine whether the conventional limitations of interference, CSI dependence, and  scalability issues remain critical. The developed analytical framework revealed an attenuated sensitivity of semantic performance to SINR, as well as performance inferiority under high-SINR regimes due to  semantic reconstruction bias. Through performance comparisons of MF and ZF under 
    both perfect and imperfect CSI cases, the results demonstrated that MF achieves semantic performance comparable to ZF.  These results suggest that semantic MIMO can relax the need for aggressive interference mitigation and highly accurate CSI, while alleviating computational and implementation burdens. 

		\bibliographystyle{IEEEtran}
		\bibliography{reference}

\begin{thebibliography}{10}
\providecommand{\url}[1]{#1}
\csname url@samestyle\endcsname
\providecommand{\newblock}{\relax}
\providecommand{\bibinfo}[2]{#2}
\providecommand{\BIBentrySTDinterwordspacing}{\spaceskip=0pt\relax}
\providecommand{\BIBentryALTinterwordstretchfactor}{4}
\providecommand{\BIBentryALTinterwordspacing}{\spaceskip=\fontdimen2\font plus
\BIBentryALTinterwordstretchfactor\fontdimen3\font minus
  \fontdimen4\font\relax}
\providecommand{\BIBforeignlanguage}[2]{{%
\expandafter\ifx\csname l@#1\endcsname\relax
\typeout{** WARNING: IEEEtran.bst: No hyphenation pattern has been}%
\typeout{** loaded for the language `#1'. Using the pattern for}%
\typeout{** the default language instead.}%
\else
\language=\csname l@#1\endcsname
\fi
#2}}
\providecommand{\BIBdecl}{\relax}
\BIBdecl

\bibitem{gunduz2022beyond}
D.~G{\"u}nd{\"u}z, Z.~Qin, I.~E. Aguerri, H.~S. Dhillon, Z.~Yang, A.~Yener,
  K.~K. Wong, and C.-B. Chae, ``Beyond transmitting bits: Context, semantics,
  and task-oriented communications,'' \emph{IEEE J. Sel. Areas Commun.},
  vol.~41, no.~1, pp. 5--41, 2022.

\bibitem{xu2025lightcom}
\BIBentryALTinterwordspacing
C.~Xu, S.~Zhang, Y.~Ma, and R.~Tafazolli, ``Lightcom: A generative
  {AI}-augmented framework for {QoE}-oriented communications,'' 2025. [Online].
  Available: \url{https://arxiv.org/abs/2507.17352}
\BIBentrySTDinterwordspacing

\bibitem{xu2026uplink}
\BIBentryALTinterwordspacing
C.~Xu, Z.~Ding, Y.~Ma, R.~Tafazolli, and P.~Zhu, ``Inference-driven uplink for
  {6G}: Architecture, principles, and challenges,'' 2026. [Online]. Available:
  \url{https://arxiv.org/abs/2508.09348}
\BIBentrySTDinterwordspacing

\bibitem{heath2018foundations}
R.~W. Heath~Jr and A.~Lozano, \emph{Foundations of {MIMO} communication}.\hskip
  1em plus 0.5em minus 0.4em\relax Cambridge University Press, 2018.

\bibitem{love2008overview}
D.~J. Love, R.~W. Heath, V.~K. Lau, D.~Gesbert, B.~D. Rao, and M.~Andrews, ``An
  overview of limited feedback in wireless communication systems,'' \emph{IEEE
  J. Sel. Area Commun.}, vol.~26, no.~8, pp. 1341--1365, 2008.

\bibitem{larsson2014massive}
E.~G. Larsson, O.~Edfors, F.~Tufvesson, and T.~L. Marzetta, ``Massive {MIMO}
  for next generation wireless systems,'' \emph{IEEE Commun. Mag.}, vol.~52,
  no.~2, pp. 186--195, 2014.

\bibitem{lu2014overview}
L.~Lu, G.~Y. Li, A.~L. Swindlehurst, A.~Ashikhmin, and R.~Zhang, ``An overview
  of massive {MIMO}: Benefits and challenges,'' \emph{IEEE J. Sel. Topics
  Signal Process.}, vol.~8, no.~5, pp. 742--758, 2014.

\bibitem{wang2022transformer}
Y.~Wang, Z.~Gao, D.~Zheng, S.~Chen, D.~G{\"u}nd{\"u}z, and H.~V. Poor,
  ``Transformer-empowered {6G} intelligent networks: From massive {MIMO}
  processing to semantic communication,'' \emph{IEEE Wireless Commun.},
  vol.~30, no.~6, pp. 127--135, 2022.

\bibitem{weng2024semantic}
Z.~Weng, Z.~Qin, H.~Xie, X.~Tao, and K.~B. Letaief, ``Semantic {MIMO} systems
  for speech-to-text transmission,'' \emph{IEEE Trans. Wireless Commun.}, 2024.

\bibitem{wu2024deep}
H.~Wu, Y.~Shao, C.~Bian, K.~Mikolajczyk, and D.~G{\"u}nd{\"u}z, ``Deep joint
  source-channel coding for adaptive image transmission over {MIMO} channels,''
  \emph{IEEE Trans. Wireless Commun.}, 2024.

\bibitem{liang2025vision}
S.~D. Liang, ``Vision language models for massive {MIMO} semantic
  communication,'' in \emph{Pro. Computer Vision Pattern Recog. Conf. (CVPR)},
  2025, pp. 1669--1679.

\bibitem{xu2025dataimportanceJ}
\BIBentryALTinterwordspacing
C.~Xu, Y.~Ma, R.~Tafazolli, and J.~Wang, ``Data-importance-aware power
  allocation for adaptive real-time communication in computer vision
  applications,'' \emph{IEEE J. S. Areas Commun. (Accepted with Minor
  Revision)}, 2025. [Online]. Available: \url{https://arxiv.org/abs/2504.08922}
\BIBentrySTDinterwordspacing

\bibitem{Zhu_2023_CVPR}
Y.~Zhu, K.~Zhang, J.~Liang, J.~Cao, B.~Wen, R.~Timofte, and L.~Van~Gool,
  ``Denoising diffusion models for plug-and-play image restoration,'' in
  \emph{IEEE/CVF Conf. Computer Vision Pattern Recognition (CVPR) Workshops},
  June 2023, pp. 1219--1229.

\bibitem{yip2007digital}
A.~M. Yip, ``Digital image restoration.'' \emph{Innovation}, vol.~7, no.~1,
  2007.

\bibitem{yu2024scaling}
F.~Yu, J.~Gu, Z.~Li, J.~Hu, X.~Kong, X.~Wang, J.~He, Y.~Qiao, and C.~Dong,
  ``Scaling up to excellence: Practicing model scaling for photo-realistic
  image restoration in the wild,'' 2024.

\end{thebibliography}
		 
	\end{document}